\documentclass[twocolumn]{aastex7}

\usepackage{amsmath}
\usepackage{booktabs}
 
 \usepackage{amsmath}

\newcommand{\pvt}{\mathrm{PVI^2}}

\newcommand{\beq}{\begin{equation}}
\newcommand{\eeq}{\end{equation}}

\newcommand{\durms}{{\delta u^{\mathrm{rms}}}}

\newcommand{\desh}{{D}_E^{\mathrm{SH}}}
\newcommand{\derms}{{D}_E^{\mathrm{rms}}}
\newcommand{\deint}{{D}_E^{\mathrm{int}}}

\newcommand{\Qsh}{{Q}_{\mathrm{SH}}}

\newcommand{\letp}{\mathrm{LET}^+}
\newcommand{\letm}{\mathrm{LET}^-}

\newcommand{\Qrms}{{Q}_\mathrm{SH}^{\mathrm{rms}}}
\newcommand{\Qint}{{Q}_\mathrm{SH}^{\mathrm{int}}}

\newcommand\kms{\mathrm{km/s}}

\begin{document}

\title{Intermittent turbulence as the driver of stochastic heating in the near-Sun solar wind}

\shorttitle{Stochastic Heating of the near-Sun Solar Wind}
\shortauthors{Ervin et al.}


\author[0000-0002-8475-8606]{Tamar Ervin}
\affiliation{Department of Physics, University of California, Berkeley, CA 94720-7300, USA}
\affiliation{Space Sciences Laboratory, University of California, Berkeley, CA 94720-7450, USA}
\email{tamarervin@berkeley.edu} 

\author[0000-0002-4625-3332]{Trevor A. Bowen}
\affiliation{Space Sciences Laboratory, University of California, Berkeley, CA 94720-7450, USA}
\email{tbowen@berkeley.edu}

\author[0000-0001-9202-1340]{Alfred Mallet}
\affiliation{Space Sciences Laboratory, University of California, Berkeley, CA 94720-7450, USA}
\email{alfred.mallet@berkeley.edu}

\author[0000-0002-1989-3596]{Stuart D. Bale}
\affiliation{Department of Physics, University of California, Berkeley, CA 94720-7300, USA}
\affiliation{Space Sciences Laboratory, University of California, Berkeley, CA 94720-7450, USA}
\affiliation{The Blackett Laboratory, Imperial College London, London, SW7 2AZ, UK}
\email{bale@berkeley.edu} 

\author[0000-0001-6038-1923]{Kristopher G. Klein}
\affiliation{Lunar and Planetary Laboratory, University of Arizona, Tucson, AZ}
\email{kgklein@arizona.edu}

\author[0000-0002-7365-0472]{Mihailo Martinović}
\affiliation{Lunar and Planetary Laboratory, University of Arizona, Tucson, AZ}
\email{mmartinovic@arizona.edu}

\author[0000-0003-4177-3328]{Benjamin D. G. Chandran}
\affiliation{Department of Physics \& Astronomy, University of New Hampshire, Durham, NH 03824, USA}
\email{benjamin.chandran@unh.edu}

\author[0000-0003-0505-8546]{Philip A. Isenberg}
\affiliation{Department of Physics \& Astronomy, University of New Hampshire, Durham, NH 03824, USA}
\email{phil.isenberg@unh.edu}

\author[0000-0002-4559-2199]{Orlando Romeo}
\affiliation{Space Sciences Laboratory, University of California, Berkeley, CA 94720-7450, USA}
\email{romeo@berkeley.edu}

\author[0000-0002-0396-0547]{Roberto Livi}
\affiliation{Space Sciences Laboratory, University of California, Berkeley, CA 94720-7450, USA}
\email{rlivi@berkeley.edu} 

\author[]{Ali Rahmati}
\affiliation{Space Sciences Laboratory, University of California, Berkeley, CA 94720-7450, USA}
\email{rahmati@berkeley.edu} 

\author[0000-0002-0396-0547]{Davin E. Larson}
\affiliation{Space Sciences Laboratory, University of California, Berkeley, CA 94720-7450, USA}
\email{davin@berkeley.edu}

\begin{abstract}
Understanding the mechanism by which the energy is dissipated at small scales to heat the outer corona and expanding solar wind, is a primary goal of the Parker Solar Probe (PSP) mission, and an outstanding question in heliophysics. Recent observational case studies have suggested that the dominant dissipation mechanism(s), stochastic heating (SH) or resonant heating (RH), may be modulated by the Alfv\'en `surface' ($M_A = 1$), while theoretical and simulation works predict that the mechanism should be modulated by the level of turbulence imbalance. 
We measure the local energy transfer (LET) rate from $9.8$ to $55$ solar radii from PSP observations, showing that the LET rate peaks at $M_A \sim 0.4 - 0.9$. We connect enhanced LET rates to signatures of intermittency, highlighting the key role that intermittent fluctuations play in heating the near-Sun solar wind. 
We compare the LET rate with estimates of the SH rate, finding that below $\sim 15$ solar radii the SH heating rate matches the LET rate provided intermittency is accounted for, regardless of $M_A$, while farther from the Sun, SH is unable to account for the observed LET rate. In contrast to predictions by recent theoretical works, this implies that `imbalanced' intermittent stochastic heating may be a viable mechanism to heat highly imbalanced, low-$\beta$ collisionless plasmas, and is likely an important mechanism of dissipation in the outer solar corona and near-Sun solar wind. 
\end{abstract}

\section{Introduction} \label{sec:intro}

Fully constraining the mechanisms through which energy is dissipated in collisionless plasmas to heat particles, has implications for heliospheric \citep[e.g.][]{Haggerty-2023, Ervin-2026natcomm} and astrophysical \citep[e.g.][]{Chael-2018, Quataert-1998, Gruzinov-1998} systems. In these systems, energy is injected at large spatial scales then undergoes a turbulent cascade to smaller scales \citep{Kolmogorov-1941, Batchelor-1953}, eventually reaching kinetic scales where dissipation often occurs \citep[e.g.][]{Howes-2008, Markovskii-2006, Bowen-2020d, Sahraoui-2010}. Kinetic-scale measurements of heliospheric plasmas, such as the solar wind, provide a means to study these processes, and apply results to systems without direct measurements but similar dimensionless parameters \citep[see][for an example of how this could be done]{Howes-2024}. 

At the Sun, the large-scale injection of energy is driven by the solar dynamo and convection on granular \citep{Bailey-2025} and super-granular \citep{Sharma-2023} scales. As the solar wind expands and propagates outwards from the solar corona, temperatures fall off more slowly than expected from adiabatic expansion \citep{Richardson-1995, Hellinger-2013, Rivera-2024}. Additionally, heavy ions are preferentially heated relative to protons \citep{Kohl-1998, Esser-1999, Tracy-2016, Rivera-2025oxygen}, and velocity distribution functions (VDF) show strong perpendicular anisotropies with larger perpendicular than parallel temperatures \citep{Marsch-1982VDF,Matteini-2007, Hellinger-2011}, in contrast to what is expected from conservation of the adiabatic invariants \citep{Chew-1956}. Observations have also shown an enhancement in the non-Maxwellian features of the VDF in plasma closer to the Sun \citep{Marsch-1982VDF, Bowen-2023-ursi}, indicative of collisionless processes at work. 

A number of dissipation mechanisms have been proposed to explain these observational signatures. Some examples are resonant heating (RH) via ion-cyclotron waves (ICWs) \citep[][and references therin]{Isenberg-2019, Cranmner-2000, Yerger-2026, Hollweg-Isenberg-2002}, stochastic heating (SH) via large-amplitude, broadband fluctuations \citep{Chandran-2010, Klein-2016, Mallet-2026}, and magnetic reconnection \citep{Hesse-2006, Sundkvist-2007, Loureiro-2017, Vech-2018, Mallet-2017}. Recent theoretical works have suggested a ``unified" phenomenology of stochastic heating and resonant heating \citep{Johnston-2025, Mallet-2026} \citep[and also magnetic reconnection in the case of][]{Mallet-2026}, where the dominant mechanism is determined by the level of turbulence imbalance, or how unequal the inward and outward propagating wave energy fluxes are (often quantified by the cross helicity), in the system. The work of \citet{Johnston-2025} suggests that the level of imbalance controls the  turbulent spectrum, such that higher levels of imbalance lead to a sharper peak in frequency space and stronger resonant heating, and thus they predict that in highly imbalanced intervals, RH should dominate while in balanced intervals, SH should dominate. Imbalanced simulations have shown evidence of quasi-linear diffusion along constant energy contours in phase space that are predicted signatures of RH, and flattening of distribution functions in balanced cases, a predicted signature of SH \citep{Zhang-2025}\footnote{In highly imbalanced turbulence, the helicity barrier causes the power spectrum to steepen to~$\sim k_\perp^{-4}$ above a perpendicular wavenumber $k_\perp^\ast$ that is substantially smaller than~$\rho_i^{-1}$. As a consequence, the fluctuations that are most effective at heating ions are at $k_\perp \sim k_\perp^\ast$, where their parallel phase velocities are very close to $v_A$. In the reference frame $R^+$ that moves anti-Sunward through the plasma at speed $v_A$ along the magnetic-field lines, the electric field $\mathbf{E}$ of the outward-propagating Alfvén waves vanishes. If these waves stochastically heat the ions, the ions will diffuse along contours of constant energy in frame~$R^+$ (because $\mathbf{E}$ vanishes in that frame), producing constant-$f$ contours that are very similar to the resonant-interaction case. This makes it difficult to distinguish whether resonant or non-resonant interactions are heating the plasma through the shape of the VDF.}. Theoretical works have derived analytical expressions for heating rates ($Q$) associated with many of these mechanisms \citep[e.g.][]{Howes-2024, Chandran-2010, Mallet-2026, Johnston-2025}, which can be quantified with spacecraft measurements and then tested against ``ground-truth" calculations of the heating rate or turbulent cascade rate \citep[e.g.][]{Politano-Pouquet-1998, Bourouaine-2013,Klein-2016-FPC, Sorriso-Valvo-2018, Ervin-2026prl}. 


Prior work has found that the quasi-linear heating rate for RH via ICWs matched, or was a significant fraction of, the turbulent cascade rate in single, imbalanced super-Alfv\'enic streams at 30 $R_\odot$ \citep{Bowen-2022} and from 14 to 55 $R_\odot$ with the ICW heating rate dropping off relative to the turbulent cascade rate at the closest distances in this radial scan \citep{Bowen-2024apjl}. This was also observed in a global study at distances of 35 to 53 $R_\odot$ in early PSP Encounters \citep{Niranjana-2024} . The presence of ICWs has also been correlated with both large-scale (e.g. cross helicity) and kinetic-scale signatures of turbulence \citep{Bowen-2024}. There is also strong evidence showing the presence of ICWs throughout the inner heliosphere \citep{He-2011, Podesta-2011, Bowen-2020d, Niranjana-2026}, associated with steepening of the turbulent spectrum \citep{Bowen-2024, McIntyre-2025, Niranjana-2026} as predicted to result from the helicity barrier in imbalanced turbulence \citep{Meyrand-2021, Squire-2022}. 

Previous work has shown that the SH rate increases closer to the Sun \citep{Martinovic-2019, Martinovic-2020}, indicating it is likely important in the sub-Alfv\'enic outer corona. In one example of the the sub-Alfv\'enic wind at $\sim 19 R_\odot$, \citet{Bowen-2025prl} showed that the RH rate was far too small to match the local energy transfer rate and found that SH, accounting for the effects of intermittency, reaches the required magnitude. \citet{Ervin-2026prl} studied the same sub-Alfv\'enic interval, making a direct measurement of the perpendicular velocity-space diffusion coefficient and velocity-space heating rate, showing that the diffusion predicted by SH was present in this sub-Alfv\'enic stream. They also found that the stochastic heating rate, including intermittency, matched the velocity-space signature of the empirical heating rate, while that of RH via ICWs did not.

Regardless of the dominant dissipation mechanism at work, intermittency is expected to enhance the local heating rate \citep{Osman-2012intheat, Mallet-2019}. Intermittent fluctuations have been shown to be prevalent throughout the inner heliosphere \citep{Osman-2012int, Cuesta-2022, Marsch-1994, Marsch-1997, Sorriso-Valvo-1999, Chhiber-2021, Osman-2014, Sioulas-2022, Bruno-2019, Matthaeus-2015, Greco-2012}, and are important for small-scale dissipation and heating \citep{Osman-2011, Osman-2012intheat, Wan-2015, Perrone-2014, Bowen-2024, Bowen-2025prl, Ervin-2026prl, Mallet-2019, Mallet-2026, Cerri-2021, Sorriso-Valvo-2018}. 

In this Letter, we show the evolution of the local energy transfer rate \citep[LET;][]{Politano-Pouquet-1998, Sorriso-Valvo-2018, Sorriso-Valvo-2015, Marsch-1997} as a function of heliocentric distance computed from Parker Solar Probe \citep[PSP;][]{Fox-2016} observations near the Sun (9.8 to 55 $R_\odot$). We find that the LET rate is larger for faster than slower solar wind at all distances, and show that enhanced LET rates are correlated with signatures of intermittency. We compare the LET rate with calculations of the SH rate ($\Qsh$) for near-Sun solar wind observed by PSP. We show a comparison with the traditional SH rate calculated using an rms spectrum of fluctuations \citep[$\Qrms$;][]{Chandran-2010, Klein-2016}, and with a SH rate that accounts for the effects of intermittency on the heating \citep[$\Qint$;][]{Bowen-2025prl, Ervin-2026prl} which is expected to have a large impact of the total heating rate \citep{Mallet-2019, Cerri-2021, Mallet-2026}, showing that the `intermittent' SH rate can account for the majority of the dissipation at distances below $\sim 15 R_\odot$, regardless of $M_A$. 


\section{Observations and Methods} \label{sec:obs}

We use near-Perihelion ($\pm$ 4 days from closest approach) particle and fields observations from PSP for Encounters 4 through 25 for this study. From Encounter 19 onward, PSP spent the majority of its time near Perihelion in the sub-Alfv\'enic wind \citep{Badman-2025, Finley-2025}. This sub-Alfv\'enic region, where $v_{sw} < v_A = |B| / \sqrt{\mu_0 \rho}$, is coupled to the Sun such that information can be transmitted in both directions, by Alfvénic fluctuations. We consider this ``Alfv\'en surface" to be the separation between the outer corona ($v_{sw} < v_A$) and super-Alfv\'enic ($v_{sw} > v_A$) solar wind. We note that there is a solar cycle bias in these observations, as PSP's perihelion distance moved closer to the Sun throughout the mission. This means that the near-Sun observations of the sub-Alfv\'enic wind all took place during the end of the rising phase and peak of solar cycle 25\footnote{Sunspot number data used to track the solar cycle from the National Oceanic and Atmospheric Administration Space Weather Prediction Center (NOAA/SWPC).}. We remove measurements that are $\pm 6$ hours of observed coronal mass ejections (CMEs) in the PSP data\footnote{CMEs were removed using a combination of the Helio4Cast catalog and prior studies. The full list can be found online \citep{Romeo-CME}.}. Rarefaction regions behind CMEs can lead to a different class of sub-Alfv\'enic regions due to a drop in plasma density enhancing the Alfv\'en speed relative to typical values at the respective distance \citep{Jagarlamudi-2025}, and thus these periods should be studied separately and we leave to future work.

We use three-axis magnetic field measurements from the FIELDS fluxgate magnetometer \citep{Bale-2016} at the native cadence (typically 0.0034 s). Partial moments of velocity and temperature from the ion electrostatic analyzer \citep[SPAN-I;][]{Livi-2022, Kasper-2016} at a 3.5-second cadence are used to calculate plasma-$\beta$, cross helicity ($\sigma_C$), residual energy ($\sigma_R$), and the Alfv\'en Mach number ($M_A = v_R / v_A$) where $v_R$ is the partial moment from PSP/SPAN-I. We apply a correction to the PSP/SPAN-I temperature tensor that accounts for systematic effects associated with the obstruction of the PSP/SPAN-I field-of-view (FOV) \citep{Badman-2025}. We note that the issues with the PSP/SPAN-I FOV has limited impact on the calculated velocity moment, but large impacts on the density moment. Thus, we use plasma density measurements ($n_e$) from quasi-thermal noise \citep[QTN;][]{Moncuquet-2020} analysis of the PSP/FIELDS radio spectrum. We assume quasi-neutrality ($n_e = n_p$) to compute the Alfv\'en speed ($v_A = |B| / \sqrt{\mu_0 \rho}$) and plasma-$\beta$ where $\beta = n_p k_B T_p / (B^2 / 2 \mu_0)$ and $T_p$ is the corrected PSP/SPAN-I temperature measurement. We note that $n_e = n_p$ ignores the effects of $\alpha$-particles, which through PSP measurements have been shown to be typically less than 4\% of the proton density \citep{Ervin-2024c, Huang-2025}. 

We filter out time periods where the PSP/SPAN-I FOV is poor, such that the majority of the ion distribution is not observed by the spacecraft and thus the plasma moments are not well measured following \citet{Romeo-Thesis, Romeo-2023}. This is done by computing the area within the FOV range in both the $\theta$ and $\phi$ directions of 1D Gaussian fits of summed counts over the deflection bins ($\theta$) and front anode bins ($\phi$) in units of differential energy flux as a function of angle. The integrated area under the 1D Gaussian fits within the FOV range of the instrument gives the FOV metrics $A_\theta$ and $A_\phi$. These are multiplied to determine a combined FOV metric $A_{\mathrm{FOV}}$. A threshold of $A_{\mathrm{FOV}} = 0.865$ is used to determine when the ion distribution is well measured by PSP/SPAN-I, and plasma moments are trustworthy. This threshold is determined from fits to the ion-to-electron density ratio ($ n_i / n_e$) as a function of $A_{\mathrm{FOV}}$ where $n_i= n_p + 2 n_\alpha$ from PSP/SPAN-I moments and $n_e$ is the QTN density \citep{Romeo-Thesis}. It is equivalent to the PSP/SPAN-I density being within 15\% of the QTN density, and is the threshold used to calibrate the PSP/SPAN-E instrument \citep{Romeo-Thesis, Whittlesey-2020}. This is a standard method of processing the PSP/SPAN-I measurements \citep[e.g.][]{Ervin-2025, Yogesh-2026, Niranjana-2026}.

We compute $\sigma_C$ and $\sigma_R$ as:
\begin{align*} \label{eqn:sigma}
        \sigma_C = \frac{\langle E^+ \rangle - \langle E^-\rangle}{\langle E^+ \rangle + \langle E^-\rangle} \\
    \sigma_R = \frac{2 \langle z^+ \cdot z^- \rangle}{\langle E^+ \rangle + \langle E^-\rangle}
\end{align*} where $\langle \cdot \cdot \cdot \rangle$ corresponds to an average over the injection timescale based on the heliocentric distance of the measurement \citep{Sioulas-2023-spectral}. $E^\pm$ is the power in the Elsasser variables, $z^\pm = v \pm v_A$ \citep{Elsasser-1950}. 

\subsection{Estimate of the Local Energy Transfer Rate}\label{sec:methods-let}

We estimate the local energy transfer (LET) rate using increments of the Elsasser variables at time lags of $\tau = 3.5$ and $49$ seconds. $3.5$ seconds is the smallest time lag possible with the PSP/SPAN-I measurements, and both time-lags corresponds to inertial range scales at PSP distances in this study \citep[see Figure 2 of][]{Sioulas-2023-spectral}. Increments are defined as $\Delta A = A(t + \tau) - A(t)$ and are used to calculate mixed third-order moments of the increments of the Elsasser variables,
\begin{equation} \label{eqn:let}
    \varepsilon^{\pm, \tau} (t) = -\frac{3}{4} \frac{|\Delta z^\pm|^2 \Delta z^\mp_\parallel}{v_{sw} \tau}
\end{equation} to estimate the LET per unit mass as a function of time: $ \varepsilon^\tau (t) = (\varepsilon^{+, \tau}(t) + \varepsilon^{-, \tau} (t)) / 2$ \citep{Marsch-1997, Sorriso-Valvo-2015, Sorriso-Valvo-2018}\footnote{This formulation of the LET rate is a first-order estimate of the energy transfer rate that assumes stationarity, isotropy, and incompressibility of the plasma turbulence \citep{Sorriso-Valvo-2018}. It neglects scale-dependent dynamic alignment which is a supported assumption near the Sun (see Figure 7b of \citet{Sioulas-2025-anisotropy} which suggests that the alignment between $z^+$ and $z^-$ is neither scale-dependent, nor particularly strong, in the near-Sun solar wind).}. $\varepsilon = \langle \varepsilon^\tau(t) \rangle_{T}$ is an average over a time window of length $T = 56$ seconds. $\Delta z^\mp_\parallel$ is a projection of the increment onto the direction of the lag. Measuring $z^-$ with PSP data, especially at small scales, is difficult \citep[for example, see Appendix B of][]{Ervin-2024b}. The FOV filtering discards intervals where the velocity distribution, and thus the Elsasser variables, are not well measured. Appendix~\ref{app:let} discusses the sensitivity of this result. In this work, we define the LET rate in units of $\mathrm{W / m^3}$ as $\mathrm{LET} = \varepsilon \rho$ where $\rho$ is the mass density. 


\subsection{Estimate of the Stochastic Heating Rate} \label{sec:methods-sh}

The calculation of the SH rate requires an estimate of the electric field fluctuations ($\delta E$) defined as $\delta E \sim \delta u B_0 / c$ \citep{Chandran-2010}, where $\delta u$ is the $E \times B$ velocity serving as a proxy for $\delta E$, not the fluctuating bulk velocity. We determine $\delta u$ as a function of scale [$\delta u(v_\perp)$] from in-situ measurements following prior work \citep{Bowen-2025prl}. We first calculate $\delta B(f)$ from PSP/FIELDS measurements using second-order structure functions of five-point increments:
\begin{equation}
\begin{split}
    \delta B_i = \Delta_{\mathrm{5 pt}} B_i =  \frac{1}{\sqrt{35}} \bigl[B_i(t - 2 \tau) - 4 B_i(t - \tau) + \\ 6 B_i(t) - 4 B_i(t + \tau) + B_i(t + 2 \tau) \bigr].
\end{split}  
\end{equation}

We relate $\delta B$ to $\delta u$ as: $\delta u = v_{\mathrm{ph}}^{\mathrm{KAW}} \delta B / B_0$ where $v_{\mathrm{ph}}^{\mathrm{KAW}} = v_A \sqrt{1 + k_\perp^2 \rho_{\mathrm{th}}^2}$ \citep{Hollweg-1999}. This relates fluctuations at $k_\perp \rho_p = 1$ to $f$ with $2 \pi f = k_\perp v_{sw} \sin{\theta_{VB}}$ which assumes Alfv\'enic turbulence and a conversion to a KAW cascade at the thermal gyroradius ($\rho_{\mathrm{th}} = v_{\mathrm{th}, \perp} / \Omega_p$ where $\Omega_p = q B / m_p$). This is applicable at the largest kinetic, sub-ion, scales where $\omega < \Omega_{p}$. 

We then use the modified Taylor hypothesis \citep{Klein-2015}, to relate fluctuations at frequency $f$ to fluctuations at a perpendicular velocity $v_\perp$ associated with each particle's gyroscale \citep{Bowen-2025prl}. The modified Taylor hypothesis (MTH) is primarily relevant in the sub-Alfv\'enic solar wind, but we note it reduces to the traditional Taylor hypothesis in the regime where $v_{SW} \gg v_A$. Assuming $k_\perp \rho_p = 1$ gives $k_\perp = \Omega_{p} / v_\perp$. We assume critically balanced turbulence \citep{Goldreich-Sridhar-1995} to relate $v_\perp$ and $f$:
\begin{equation} \label{eqn:vperp}
     v_\perp = \frac{\Omega_p}{2 \pi f} \left[ \durms (f) \left(1 + \frac{v_{SW}}{v_A} \cos{\theta_{\mathrm{VB}}} \right) 
         +  v_{SW} \sin{\theta_{\mathrm{VB}}} \right] 
\end{equation}
where $\durms (f)$ is the rms fluctuation spectrum defined as $\durms(f) = \sqrt{\langle \delta u(f, t)^2 \rangle_{T} }$. We note that prior work has also shown that this held for a purely perpendicular ($k_\parallel = 0$) 2D spectrum \citep{Bowen-2025prl}. We only look at time periods where the flow angle relative to the background field ($\theta_{\mathrm{VB}}$) is $\sim$ perpendicular ($70 < \theta_{\mathrm{VB}} < 110$), to look at the perpendicular spectrum.

Using the MTH, we determine $\delta u(v_\perp)$ from $\delta u(f)$, from which we compute energy diffusion coefficients that are then used to compute the stochastic heating rate \citep{Chandran-2010}. The energy diffusion coefficient is defined in a scale-dependent manner \citep{Bowen-2025prl}:
\begin{equation}
    \desh(v_\perp) = c_1 v_\perp \Omega_p \delta u^3(v_\perp) e^{-c_2 / \epsilon}
\end{equation} \label{eqn:deperp-sh} where $\epsilon = \delta u(v_\perp) / v_\perp$ is the ``stochasticity parameter". $v_\perp$ is defined from Eqn.~\ref{eqn:vperp} using the MTH based on $v_{SW}$, $\theta_{\mathrm{VB}}$ and $v_A$ over the interval $T$. $c_1$ and $c_2$ are order unity constants set to 0.75 and 0.34 respectively \citep{Chandran-2010} and have been shown to be consistent with PSP observations in the near-Sun solar wind \citep{Bowen-2025prl, Ervin-2026prl}. At scales below $\rho_{\mathrm{th}}$, the scaling of the electric field fluctuation spectrum changes. However, it was shown through an empirical measurement of the velocity-space heating rate in a sub-Alfv\'enic stream that the majority of the in-situ heating \citep[see Figure 3 of][]{Ervin-2026prl}, and the SH \citep[see Figure 3 of][]{Bowen-2025prl}, occurs at larger scales, $\rho_p > \rho_{\mathrm{th}}$, where the KAW dispersion relation holds.


We define a `rms' ($\derms$) and `intermittent' ($\deint$) energy diffusion coefficient. $\derms$ is the ``traditional" coefficient \citep{Chandran-2010} which uses the rms value of the fluctuations $\durms(v_\perp) = \sqrt{\langle \delta u(v_\perp, t)^2 \rangle_{T} }$. This averages out infrequent, but large amplitude fluctuations. $\deint$ accounts for the effects of intermittency on the heating, by treating $\delta u$ as a random variable: $\deint = \langle | \desh (v_\perp, t) | \rangle_{T}$  \citep{Bowen-2025prl}. 

From $\derms$ and $\deint$, rms ($\Qrms$) and intermittent ($\Qint$) SH rates for the formulation applicable at $\beta < 1$ are calculated as \citep{Chandran-2010}:
\begin{equation} \label{eqn:Qsh}
    \Qsh = \int d \Qsh = -2 \pi m_i \int_0^\infty \desh (v_\perp) \frac{\partial f_\perp}{\partial v_\perp} dv_\perp .
\end{equation} 
To compute $\partial f_\perp / \partial v_\perp$ in Eqn.~\ref{eqn:Qsh}, we define a Maxwellian 1D VDF $f_\perp(v_\perp)$ on the $v_\perp$ grid, normalized to the measured QTN density. Using the observed distribution functions for a statistical study of this size is unfeasible. In Appendix~\ref{app:sh-vdf} we discuss the sensitivity of this calculation to departures from Maxwellian distributions, specifically flattening of the distribution at low $v_\perp$ to replicate the flat-top distribution functions shown to be produced by SH \citep{Klein-2016}. As in the calculation of the LET, $T = 56$ seconds is used to compute $\Qrms$ and $\Qint$.


\subsection{Signatures of Intermittency}  \label{sec:methods-intermittency}
We measure the partial variance of increments (PVI), which identifies intermittent structures by evaluating gradients at a specifc scale, including both magnetic field ($\Delta \mathbf{B}$) and velocity ($\Delta \mathbf{v}$) increments \citep{Sorriso-Valvo-2018}:
\begin{equation}
    \mathrm{PVI^2}_\tau(t) = \frac{\Delta \mathbf{B}(\tau, t)^2}{\langle \Delta \mathbf{B}(\tau, t)^2 \rangle} + \frac{\Delta \mathbf{v}(\tau, t)^2}{\langle \Delta \mathbf{v}(\tau, t)^2 \rangle}.
\end{equation} $\Delta \mathbf{B}(t)$ and $\Delta \mathbf{v}(t)$ are vector increments computed at two time lags ($\tau = 3.5$ and $49$ s) as in the estimate of the LET\footnote{The PVI measurement can also be made using only $\Delta B$. We use this definition with both $\Delta B$ and $\Delta v$, as the SH rate depends on the electric field and $\mathbf{E} = - \mathbf{v} \times \mathbf{B}$.}. $\langle \cdot \cdot \cdot \rangle$ is the ensemble average of the increments at lag $\tau$ calculated over an interval that is $5 \lambda_C$ \citep{Servidio-2011}. $\lambda_C$ is the correlation length at the mean radius of the increment, determined from statistical work on the scaling of $\lambda_C$ as a function of heliocentric distance \citep{Cuesta-2022correlation, Cuesta-2022intermittency}.

We measure the kurtosis ($\kappa$) to determine deviations from Gaussianity, defined as: $\kappa_i(\tau) = S_i^4(\tau) / S_i^2(\tau)^2$ where $S_i^p(\tau) = \langle |\Delta B_i^\tau|^p \rangle$ is the structure function of order $p$. $\langle \cdot \cdot \cdot \rangle$ is an average over an interval of observations ($N$) such that moments up to $p = 4$ are well measured. We expect to measure moments up to $p < \mathrm{log_{10}}(N) - 1$ \citep{DdW-2004}, meaning $N = 150,000$ observations is required. For the high cadence MAG data (0.0034 seconds), a 510-second interval with $N = 150,000$ observations is used. For time periods when the cadence is 0.0068 seconds, intervals of 1020-seconds are used. Non-Gaussian structure, or heavy-tailed distributions, correspond to kurtosis values of $\kappa > 3$ \citep{Marsch-1994, Sorriso-Valvo-1999}. 
We calculate the kurtosis at time lags corresponding to $1 \times 10^{-3} \leq k d_i \leq 3 \times 10^{-2}$, constraining the inertial range \citep{Sioulas-2023-spectral}. We use a modified form of the Taylor hypothesis $k = 2 \pi f / (|\mathbf{v}_{SW} + \mathbf{v}_A - \mathbf{v}_{SC}|)$ to relate spatial scales to time lags ($f = 1 / 2 \tau$) \citep{Klein-2015}, where $\mathbf{v}_{SW}$, $\mathbf{v}_A$, $\mathbf{v}_{SC}$, and $d_i = \rho_{\mathrm{th}} / \sqrt{\beta}$ are averages of the solar wind velocity, Alfv\'en speed, spacecraft velocity, and ion inertial length over the interval within which the kurtosis was calculated.



\section{Results} \label{sec:res}
\begin{figure}
    \centering
    \includegraphics[width=\linewidth]{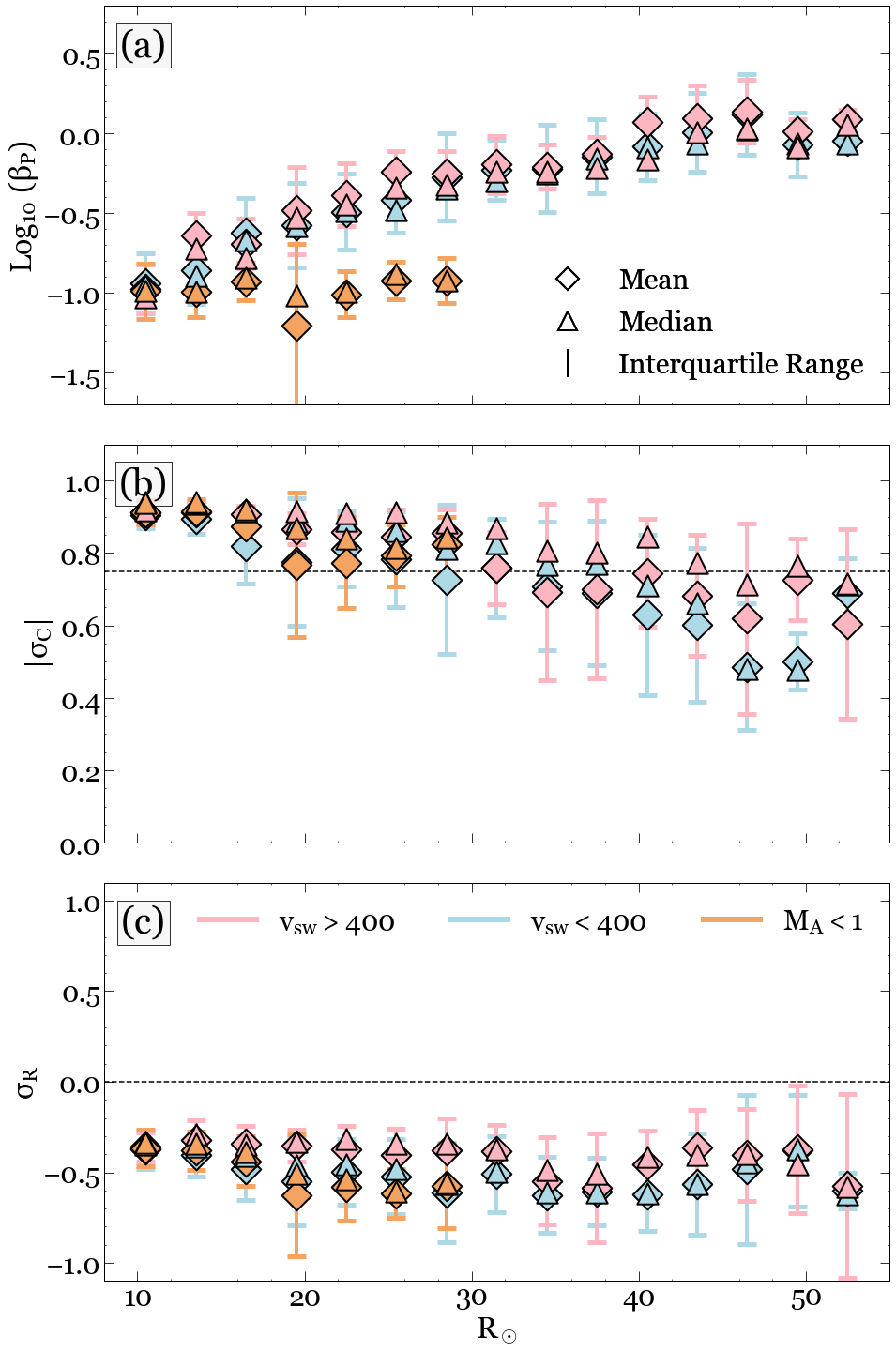}  
    \caption{(a): $\mathrm{log_{10}} \beta_\mathrm{P}$, (b): $|\sigma_C|$, and (c): $\sigma_R$ as a function of heliocentric distance in solar radii ($R_\odot$). Diamonds (triangles) show the mean (median) value in each distance bin. Fast wind ($v_{sw} > 400 \; \kms$) is shown in pink and slow wind ($v_{sw} < 400 \; \kms$) in blue. Sub-Alfv\'enic wind ($M_A < 1$) is shown in brown. In bins below 20 $R_\odot$ where the blue markers are not visible, they are covered by the brown markers. Data is in bins of 3 $R_\odot$ for calculation of the mean, median, and interquartile range.}
    \label{fig:params}
\end{figure}

Figure~\ref{fig:params} shows (a) plasma-$\beta$, (b) $|\sigma_C|$, and (c) $\sigma_R$ as a function of heliocentric distance. The computation is binned into radial distance bins of 3 $R_\odot$ showing the mean (diamond), median (triangle) and interquartile range (IQR) over each interval. We split the wind into `fast' ($v_{sw} > 400 \; {\kms}$) and `slow' ($v_{sw} < 400  \; {\kms}$) for each of these bins. Sub-Alfv\'enic wind ($M_A < 1$) is shown in brown, and is a combination of both `fast' and `slow' sub-Alfv\'enic wind. We note that choosing a threshold of 400 ${\kms}$ is arbitrary, and that the wind is still accelerating at these distances such that a constant threshold does not distinguish between wind that would be seen as characteristically `fast' or `slow' at 1 AU. At all distances, there is very little difference in parameters between the `fast' and `slow' wind, regardless of the velocity threshold chosen. The near-Sun wind is low-$\beta$, meaning the low-$\beta$ formulations of both the RH and SH mechanisms are applicable. The wind is highly imbalanced ($|\sigma_C| \sim 1$), with primarily negative residual energy meaning the plasma is magnetically dominated.

\begin{figure}
    \centering
    \includegraphics[width=\linewidth]{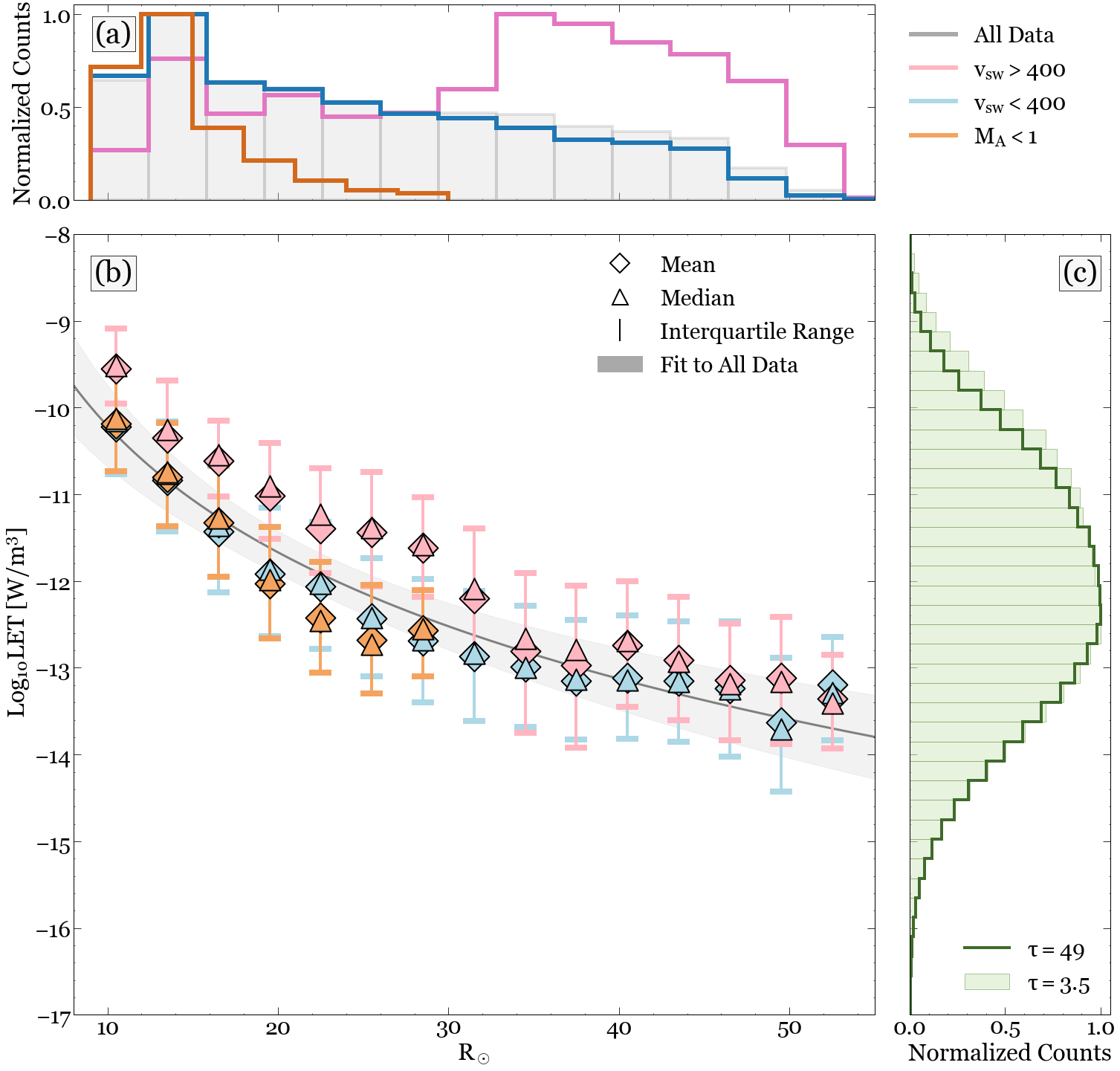}  
    \caption{(a): Histogram showing the number of calculations in each distance based bin. The distribution of all data is shown in the filled (gray) histogram. (b): LET rate ($\epsilon$) in $\mathrm{W / m^3}$ as a function of heliocentric distance ($R_\odot$) for $\tau = 3.5$ s. Diamonds (triangles) show the mean (median) value in each distance bin. Fast wind ($v_{sw} > 400 \; \kms$) is shown in pink and slow wind $v_{sw} < 400 \; \kms$ in blue. Sub-Alfv\'enic wind ($M_A < 1$) is shown in brown. In bins below 20 $R_\odot$ where the blue markers are not visible, they are covered by the brown markers. Data is in bins of 3 $R_\odot$. (c): Distribution of LET values in  $\mathrm{W / m^3}$ for two different time lags. $\tau = 3.5$ s is the filled histogram and $\tau = 49$ s is unfilled.}
    \label{fig:let}
\end{figure}

In Figure~\ref{fig:let} we show the evolution of the LET rate with heliocentric distance. Figure~\ref{fig:let}(a) shows the number of calculations of the LET rate in each radial distance bin. Figure~\ref{fig:let}(b) shows the mean, median, and interquartile range of the LET rate as a function of $R_\odot$, calculated in $\mathrm{Log_{10}}$ space. The computation is binned into radial distance bins of 3 $R_\odot$, identical to in Figure~\ref{fig:params}. These values are consistent with previous measurements of the LET rate at similar heliocentric distances \citep{Bandyopadhyay-2020, Bowen-2024apjl, Bowen-2025prl}. The distribution of LET measurements for both time lags is almost identical (panel (c)), and we use the $\tau = 3.5$ seconds time lag in all figures moving forward. Intervals with negative estimated values of the LET rate have been removed, and no absolute value of $\varepsilon$ is taken.

Figure~\ref{fig:let}(a) shows that more `slow' wind is present at close distances and more `fast' wind further from the Sun. This is to be expected, as 400 ${\kms}$ is an arbitrary velocity threshold and does not adequately account for the continued acceleration of the wind at near-Sun distances.

In Figure~\ref{fig:let}(b), the LET rate ($\tau = 3.5$ seconds) falls off as a function of heliocentric distance and is larger for faster than slower wind across almost all heliocentric distances. As stated above, 400 ${\kms}$ is an arbitrary velocity threshold, however the results that faster wind shows larger LET rates than slower wind over heliocentric distance holds regardless of how the `fast'-`slow' split is defined, provided sensible values are used. 

We do a weighted linear fit in log-log space of the LET rate as a function of $R_\odot$, as the distribution of LET values is log-normal. The weighting scheme is such that each radial distance bin is equally weighted, regardless of sample number, in the fit. This ensures that closer Sun distances do not dominate the fit due to more samples in this regime. We parameterize the LET as: $\mathrm{LET}(r) = A \left( \frac{r}{r_0} \right)^\alpha$ where $r_0 = 25 \; R_\odot$ and $r$ is in solar radii ($R_\odot$). $r_0 = 25 \; R_\odot$ is chosen as the measurements of the LET rate are well constrained there by observations. The fit is shown by the darkgray shading in Figure~\ref{fig:let}(b) and has parameters: $A = 7.27 \pm 4.03 \times 10^{-13}$ and $\alpha = -4.84 \pm 1.12$.

\begin{figure}
    \centering
    \includegraphics[width=\linewidth]{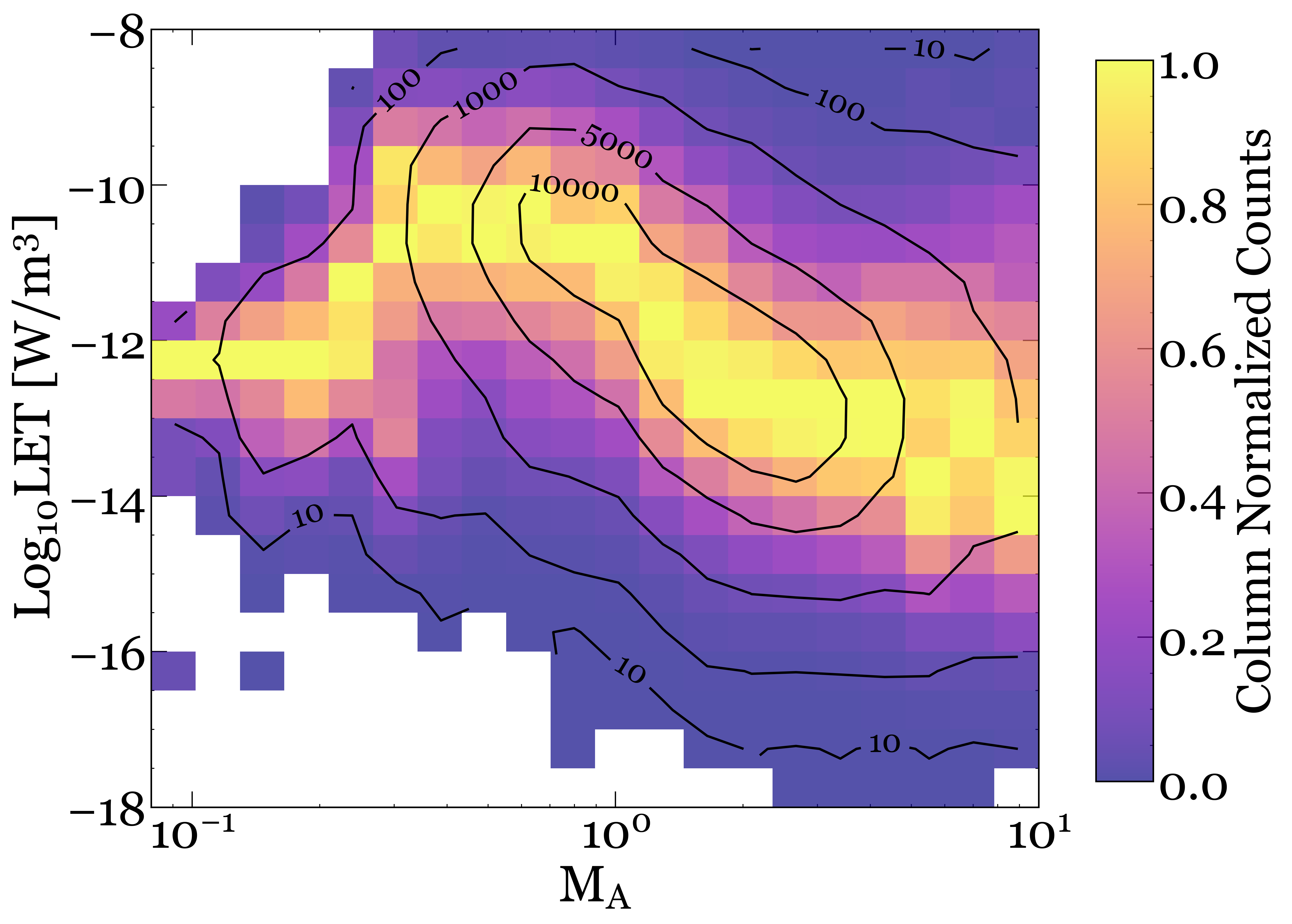}  
    \caption{Column normalized 2D histogram of $\mathrm{Log_{10} LET}$ rate ($\mathrm{W / m^3}$) as a function of $M_A$. The black lines are constant value contours in counts of the un-normalized joint distribution.}
    \label{fig:let-ma}
\end{figure}

Figure~\ref{fig:let-ma} shows a column normalized 2D histogram of $M_A$ and the LET rate. There is a peak in the local LET rate at $M_A \sim 0.4 - 0.9$. We note that there is not a direct relation between $M_A$ and heliocentric distance from the Sun, thus the trend in LET as a function of $M_A$ cannot be directly compared with Figure~\ref{fig:let}.  Only $\sim 2 \%$ of the measurements in this Figure fall below $M_A \sim 0.4$. Additional measurements in the PSP extended mission would help identify if this trend holds.

\begin{figure}
    \centering
    \includegraphics[width=\linewidth]{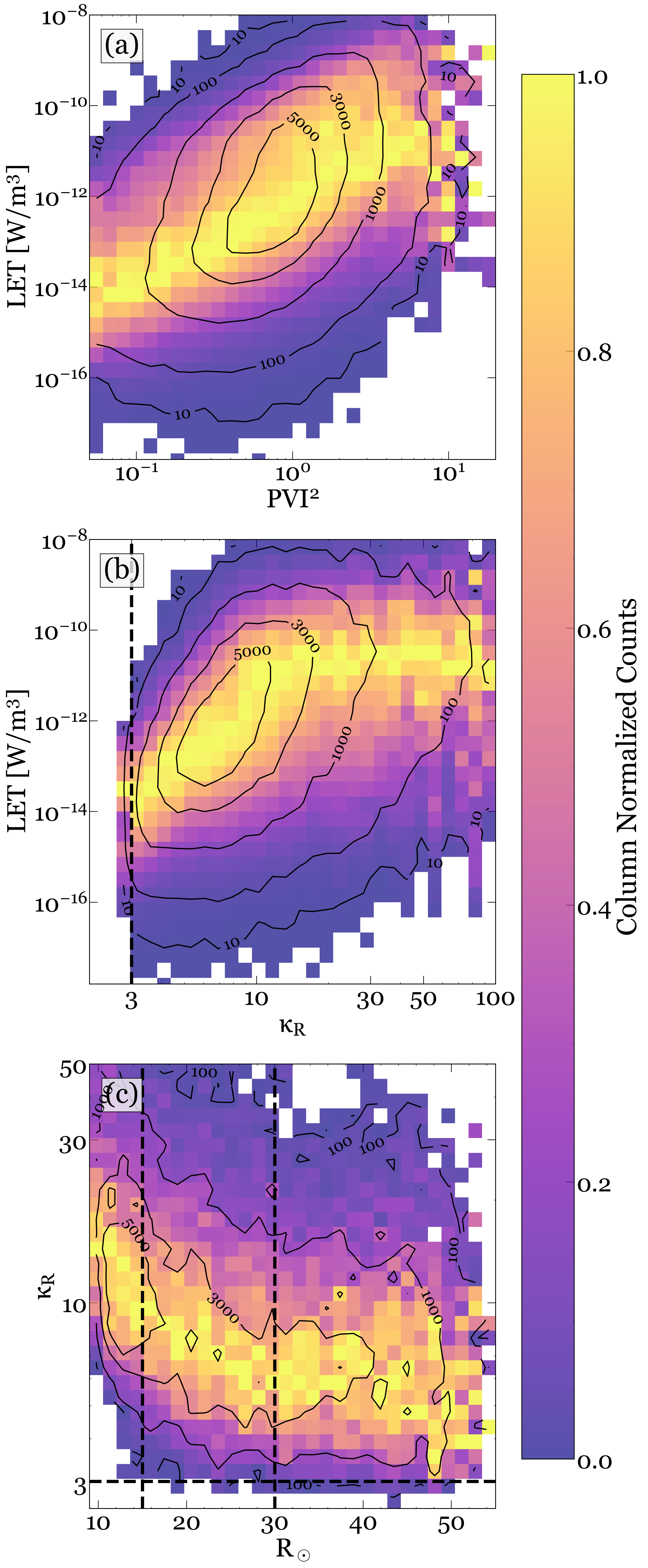}  
    \caption{Column normalized 2D histograms. The black lines are constant value contours in counts of the un-normalized joint distribution. (a): Distribution of the LET rate and $\pvt$.
    (b): Distribution of the LET rate and kurtosis of $B_R$ ($\kappa_R$). The vertical dashed line is at $\kappa_R = 3$. (c): Distribution of $\kappa_R$ as a function of heliocentric distance ($R_\odot$). The horizontal dashed line is at $\kappa_R = 3$ and vertical dashed lines at 15 and 30 $R_\odot$.}
    \label{fig:pvi-kurt-let}
\end{figure}

Figure~\ref{fig:pvi-kurt-let} relates the LET rate with signatures of intermittency using column normalized 2D histograms. The overlaid black lines show constant value contours (in counts) of the un-normalized joint distribution in each panel. Figure~\ref{fig:pvi-kurt-let}(a) is the LET rate as a function of $\pvt$. As the measure of $\pvt$ increases, so does the value of the LET rate with a correlation coefficient of 0.43. We note that above $\pvt$ of 7 and below $\pvt$ of 0.1, there are limited counts, which is likely the reason for the variance in the LET rate in these regions.

Figure~\ref{fig:pvi-kurt-let}(b) shows the LET rate as a function of $\kappa_{R}$, which is the kurtosis of the radial magnetic field. Panels (b) and (c) show the kurtosis calculation at $k d_i = 0.1$, however we note these same results hold for all lags within $1 \times 10^{-3} \leq k d_i \leq 3 \times 10^{-2}$ (see Section~\ref{sec:methods-intermittency}). For each measurement of the LET rate, we assign a $\kappa_R$ value based on the time interval the LET rate was measured in. It is important to note that these measurements are not independent, as the LET rate and $\kappa$ are both calculated with moments of the increments of the magnetic field. As $\kappa_R$ increases, so too does the LET rate. We note that there are limited counts at large values of $\kappa_R$ and for LET rates larger than $10^{-10} \; \mathrm{W / m^3}$.


Figure~\ref{fig:pvi-kurt-let}(c), shows a column normalized 2D histogram of $\kappa_{R}$ as a function of heliocentric distance. $\kappa_R$ is larger closer to the Sun, with an increase beginning at $\sim$ 30 $R_\odot$ and a more dramatic increase at $\sim$ 15 $R_\odot$.

\begin{figure}
    \centering
    \includegraphics[width=\linewidth]{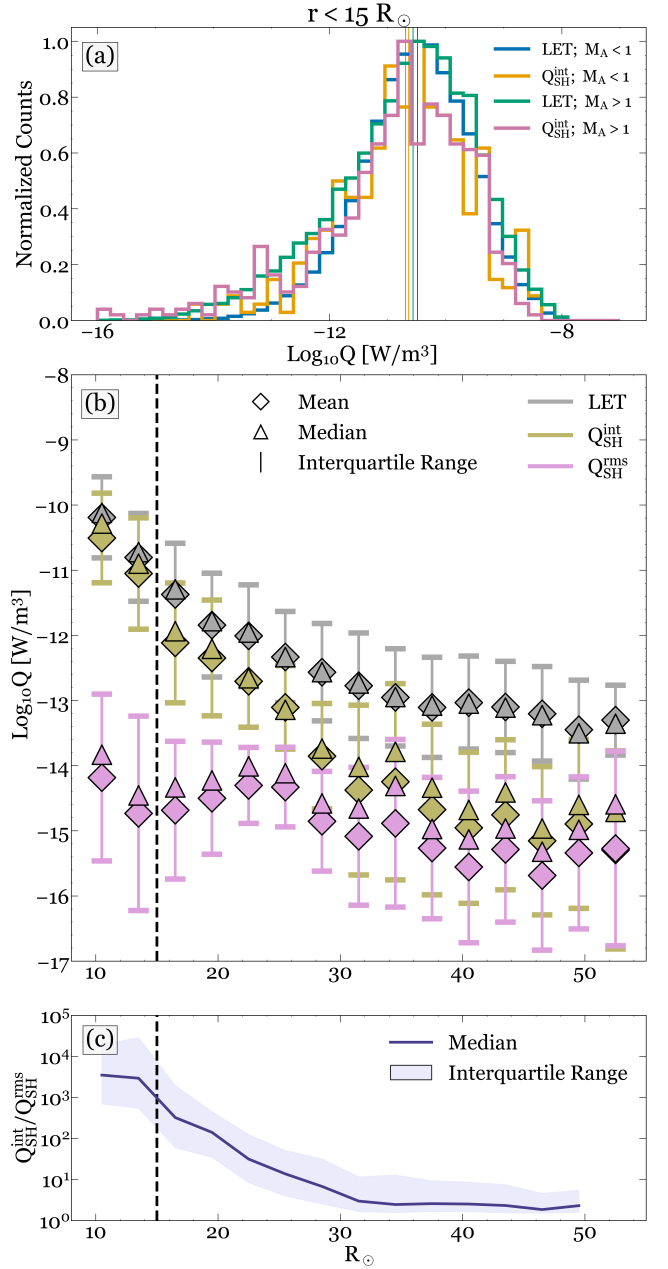}  
    \caption{Comparison of measured LET rate with computed $\Qint$ and $\Qrms$. (a): 1D histograms of LET rate and $\Qint$ below 15 $R_\odot$ for $M_A < 1$ and $M_A > 1$. Vertical lines show the median of the distribution. (b): The mean, median, and interquartile range of the LET rate (gray), $\Qint$ (olive), $\Qrms$ (purple) as a function of heliocentric distance in bins of 3 $R_\odot$. (c): $\Qint / \Qrms$ as a function of heliocentric distances. The solid line shows the median and the shaded region shows the interquartile range, calculated in bins of 3 $R_\odot$. The vertical dashed lines in (b) and (c) are at 15 $R_\odot$.}
    \label{fig:sh-let}
\end{figure}

Figure~\ref{fig:sh-let} compares the computed SH rates ($\Qint$ and $\Qrms$) with the LET rate as a function of heliocentric distance. We filter for times where the PSP/SPAN-I FOV is within requirements (as with the LET rate) and $\theta_{\mathrm{VB}} \sim 90^\circ$ (between $70^\circ$ and $110^\circ$) such that we observe perpendicular spectra. These constraints gives 7831, 56-second intervals where the SH rate was calculated. 

Figure~\ref{fig:sh-let}(a) shows 1D histograms of LET and $\Qint$ measurements below 15 $R_\odot$ for $M_A < 1$ and $M_A > 1$. The distribution of $\Qint$ and LET values are overlapping for both the $M_A > 1$ and $M_A < 1$ cases, indicating that $\Qsh \approx$ LET rate below $\sim 15 R_\odot$ regardless of $M_A$. 

Figure~\ref{fig:sh-let}(b) shows the evolution of $\Qint$ (olive), $\Qrms$ (purple), compared with the LET rate (gray) as a function of heliocentric distance. The mean, median, and interquartile range are calculated in log-space in bins of 3 $R_\odot$ (as in prior figures). Panel (c) shows the $\Qint / \Qrms$ ratio as a function of $R_\odot$ where the median is in purple, and the shaded region is the interquartile range both calculated in bins of 3~$R_\odot$. 

At all distances, $\Qrms$ is too small to match the LET rate. Including the effects of intermittency in the calculation of $\Qsh$ increases the amplitude of the SH rate significantly, between 10 - $\mathrm{10^5}$ times depending on the interval. At distances farther than $\sim$ 30 $R_\odot$, the $\Qint / \Qrms$ ratio is flat at $\sim 2$, and $\Qint$ is significantly smaller than the LET rate.  $\Qint / \Qrms$ starts to increase significantly at distances within $\sim$ 30 $R_\odot$, corresponding to the increase in intermittency (see Figure~\ref{fig:pvi-kurt-let}(c)), where the intermittent SH rate grows such that $\Qint$/LET $\sim 0.1$. At 15 $R_\odot$,the $\Qint / \Qrms$ ratio reaches values of $\sim \mathrm{10^3}$ and the intermittent SH rate is approximately the LET rate (see also panel (a)).\footnote{This result holds for 1d VDFs ($f_\perp$) that have a flatter core (are platykurtic) than a Maxwellian distribution (see Appendix~\ref{app:sh-vdf}).}

\section{Discussion} \label{sec:disc}

A primary goal of PSP is to constrain the dissipation processes in the outer corona and near-Sun solar wind \citep{Fox-2016}. We show the evolution of the local energy transfer (LET) rate as a function of heliocentric distance from 9.8 to 55 $R_\odot$, noting there can be a large variation in amplitude, even within a small range of heliocentric distance. We show that within 15 $R_\odot$, stochastic heating that includes intermittent effects can account for the observed heating, regardless of Alfv\'en Mach number. To constrain the applicability of SH, we compare the LET with computed stochastic heating rates. We show that the `rms' stochastic heating rate \citep[$\Qrms$;][]{Chandran-2010, Klein-2016} is orders of magnitude too small to account for the observed heating. The inclusion of intermittency in the SH calculation \citep[$\Qint$;][]{Bowen-2025prl} dramatically enhances the stochastic heating rate; $\Qint$ is a factor 10 - $\mathrm{10^5}$ higher than $\Qrms$, increasing closer to the Sun. Below $\sim 15 R_\odot$, $\Qint$ is typically $\mathrm{10^3}$ times larger than $\Qrms$ (if not more), and $\Qint \approx$ LET, irrespective of $M_A$. This implies that the dominance of stochastic heating is not modulated by the Alfv\'en Mach number, but is governed by the amplitude and intermittency of the turbulent fluctuations near ion kinetic scales which enable sufficient $\delta u / v_\perp \sim 1$ which results in $\Qsh \approx $ LET.


At all heliocentric distances, faster wind has larger LET rates than slower wind, consistent with the idea that wind is fast when a smaller fraction of the wave power is dissipated closer to the Sun (below the sonic point) \citep{Leer-Holzer-1980-sonicpoint, Chandran-2021}, making wave amplitudes larger in fast wind at further distances such as those observed by PSP. The LET rate peaks below $M_A = 1$, at $M_A \sim 0.4 - 0.9$, consistent with work showing a peak in the fluctuation amplitude and volumetric heating rates at these Alfv\'en Mach numbers \citep{Sioulas-2025-arxiv}. This could also support the predictions of \citet{Kasper-Klein-2019}, regarding a zone of `preferential heating' near the Alfven surface. Additional observations below the Alfv\'en surface would allow for further testing of these theoretical predictions. We note that there is not a one-to-one mapping of $\mathrm{LET}(M_A)$ to $\mathrm{LET}(r)$, meaning that streams with low $M_A$ at close distances (10 - 20 $R_\odot$) could correspond to asymptotically slow wind streams that are either underpowered in wave amplitude (e.g. from pseudostreamers) or the wave power was strongly dissipated inside the sonic critical point \citep{Leer-Holzer-1980-sonicpoint, Chandran-2021}. 

The LET rate is correlated with signatures of intermittency (PVI and kurtosis) indicating that intermittent fluctuations play an important role in heating of the near-Sun wind and outer corona \citep{Mallet-2019, Osman-2012intheat, Perrone-2014, Cuesta-2022intermittency, Wan-2015, Bowen-2025prl, Ervin-2026prl}. The role of intermittent fluctuations is thus vital to be included in large-scale models of the heliosphere to accurately model the solar wind's evolution \citep{Ervin-2026natcomm, Cranmer-2007, Chandran-2025-twofluid, Chandran-2025-intermittency} and improve space weather forecasts \citep{Waters-2026}. 

The dominant dissipation process is expected to be controlled by plasma parameters such as $\beta$ \citep{Howes-2024}, the level of imbalance (cross helicity; $\sigma_C$) \citep{Zhang-2025, Johnston-2025, Mallet-2026}, and the shape of the turbulent spectrum. The shape of the turbulent spectrum has been shown to be related to the cross helicity \citep{Podesta-2010, McIntyre-2025, Bowen-2024}, residual energy ($\sigma_R$) \citep{Bowen-2018, Ervin-2025}, and distance from the Sun \citep{Chen-2020, Sioulas-2023-spectral}. We show that near the Sun, plasma-$\beta$ is low and increases with heliocentric distance, thus we expect the low-$\beta$ formulation for both RH and SH \citep{Chandran-2010} mechanisms to be applicable.\footnote{\citet{Hoppock-2018} showed a generalization of SH at higher-$\beta$, which could be applied in other plasma regimes. For RH, approximating the wave frequency from the cold-plasma dispersion becomes more difficult, but RH can still occur at larger $\beta$.} The plasma has negative residual energy, and is highly imbalanced ($|\sigma_C| \sim 1$), especially near the Sun, and thus recent theoretical works predict that RH would be dominant \citep{Johnston-2025}.

These results, alongside prior works \citep{Bowen-2025prl, Ervin-2026prl}, show that ``imbalanced intermittent stochastic heating" may exist, indicating that the heating is mostly done in sporadic, large-amplitude intermittent events instead of continuously. This implies that any comprehensive theory, as well as numerical simulations, must have an adequate description of the statistical properties of the intermittency. Theoretical and simulation work is needed to understand if and when this kind of heating is relevant. Future work in balanced solar wind intervals is also needed to understand if SH is dominant in those time periods. Perhaps the turbulence in simulations is not intermittent enough to replicate the conditions at which SH becomes important in the near-Sun solar wind; as the intermittency strength at ion scales depends on length of the inertial range, and there are inherent limitations on the length of the inertial range due to computational constraints.

While these results provide a constraint on the magnitude and mechanism of dissipation in the near-Sun wind, they rely on bulk measurements of the heating rate and the local energy transfer rate. Measurements of the LET rate can have large uncertainties, primarily due to assumptions needed to make this measurement from a single spacecraft. Other works have used an `empirical' method to determine total proton heating rate based on the radial profile of the proton temperature \citep{Bourouaine-2013} which could be compared with estimates of the LET rate in future studies. Alternative methods, such as those measuring heating rates directly from VDFs \citep{Ervin-2026prl} or through techniques such as field-particle correlations \citep[FPC;][]{Klein-2016-FPC} are more robust. Different theoretical works have different analytic descriptions of the SH rate \citep{Chandran-2010, Klein-2016, Johnston-2025, Mallet-2026}, and varying definitions of the relevant fluctuation ($\delta u$) spectrum or scale to use for computation of the heating rate. Comparisons of the bulk heating rate cannot definitively distinguish between these analytic descriptions when predicted amplitudes are similar, meaning measurements of the velocity-space heating rate are essential for determination of the ``correct" analytical formulation \citep{Ervin-2026prl}. This work used the $\Qint$ formulation from \citet{Bowen-2025prl}, shown to have a consistent velocity-space signature when compared with empirical measurements of heating rates directly from the VDF \citep{Ervin-2026prl}. We note that these bulk SH rates strongly rely on the choice of coefficients $c_1$ and $c_2$, making the comparison to a total heating energy budget difficult, but radial trends are not dependent on $c_1$ and $c_2$ as these are controlled by fluctuation coupling and adiabatic invariant physics, respectively. Future work could fit for $c_1$ and $c_2$ to constrain these parameters \citep[e.g.][]{Xia-2013, Martinovic-2020, Bourouaine-2013}.

We also note that this Letter does not include a comparison with the quasi-linear heating rate for the resonant interaction with ICWs as has previously been done for case studies of PSP observations \citep{Bowen-2022, Bowen-2024apjl, Bowen-2025prl, Ervin-2026prl}. This heating rate requires measurements of the VDF, which can be difficult to use on a statistical basis with PSP, due to issues associated with the instrumental field-of-view, among other things. Calculation of a ``proxy" ICW heating rate could be done on a statistical basis in the future using magnetic field observations. 

These results show that intermittent fluctuations largely drive the heating in the near-Sun solar wind, and that `intermittent' stochastic heating, likely accounts for a large amount of the dissipation of turbulent fluctuations near the Sun. This provides constraints on heating in the outer corona and near-Sun solar wind, and insight into dissipation in low-$\beta$, highly imbalanced, collisionless plasmas.

\section{Acknowledgments} \label{sec:acknowledgments}
The FIELDS and SWEAP experiments on the Parker spacecraft was designed and developed under NASA contract NNN06AA01C. We acknowledge the NASA Parker Solar Probe Mission, the FIELDS team led by S. D. Bale, and the SWEAP team led by M. Stevens for use of data. 

TE acknowledges funding from The Chuck Lorre Family Foundation Big Bang Theory Graduate Fellowship, and from the Royal Astronomical Society (RAS) to attend the Solar Wind 17 meeting, from which fruitful conversations led to the advancement of this work. TAB acknowledges support from NASA Grant 80NSSC24K0272 through the HSR program. AM acknowledges support from NASA grant 80NSSC20K1284. KGK acknowledges partial support from NASA grant 80NSSC24K0171. BC acknowledges support from NASA Grant 80NSSC24K0171 and DOE grant DE-SC0026201. 

The authors thank the Referee for their suggestions that greatly improved the uncertainty analysis and quality of the manuscript.

\bibliography{ms}{}
\bibliographystyle{aasjournalv7}

\appendix{}\label{sec:appy}
\section{Sensitivity of Local Energy Transfer Rate to $z^-$} \label{app:let}

\begin{figure}[h!]
    \centering
    \includegraphics[width=0.8\linewidth]{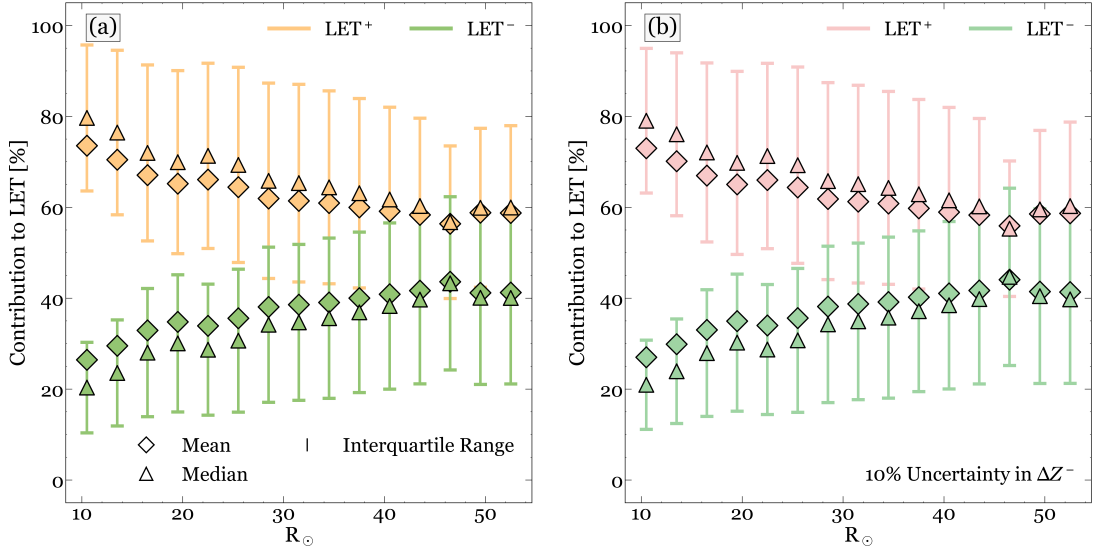}
    \caption{(a): Relative contribution of $\letp$ (orange) and $\letm$ (green) to the total LET rate. Calculated as $\mathrm{LET}^{\pm} \; \% = \mathrm{LET}^{\pm} / (\letp + \letm) \times 100 \%$. (b): Relative contribution of $\letp$ and $\letm$ with 10\% uncertainty added to the $\Delta z^-$ measurement when calculating the LET rate.}
    \label{fig:let_uncertainty}
\end{figure}

In Figure~\ref{fig:let_uncertainty}(a) we show the relative contribution of $\letp$ and $\letm$ as a function of heliocentric distance (in bins of 3~$R_\odot$) for the LET results in Figure~\ref{fig:let} and~\ref{fig:let-ma}. The reported interquartile range is not a measurement of ``uncertainty'', but rather the spread in contribution percentage within the 3~$R_\odot$ bin. The $\letm$ term contributes more at further heliocentric distances, where the cross helicity decreases (see Figure~\ref{fig:params}). 

The measurement of the local energy transfer (LET) rate, relies on increments of the Elsasser variables $\Delta z^\pm$. There are likely uncertainties associated with the $\Delta z^-$ estimate from PSP observations. While the noise in the magnetic field measurement is quite small, there is likely noise in the velocity fluctuation measurement. We expect uncertainties in $\Delta z^-$ to more strongly contribute to $\letm$ than $\letp$, as $\letm \sim |\Delta z^-|^2$ while $\letp \sim \Delta z_\parallel^-$ (see Equation~\ref{eqn:let}), and the aligned portion ($\Delta z_\parallel^-$) should be well constrained. We test the impact of noise by defining $\Delta z^- = \Delta \Tilde{z}^- + \eta^-$ where $\eta^-$ is the noise. We test reasonable noise levels (1\%, 5\%, and 10\%) relative to the $\Delta z^-$ measurement, and propagate this through the LET measurement such that it impacts both the $\letp$ and $\letm$ measurements. We run  300 iterations for each interval over which the LET rate was calculated, adding noise to the $\Delta z^-$ measurement sampled from a normal distribution.
Figure~\ref{fig:let_uncertainty}(b) shows the relative contribution of $\letp$ and $\letm$ to the total cascade rate with 10\% noise in $\Delta z^-$. We find that there is very little impact to the relative contributions and that the results of this study hold, even with reasonable uncertainty. Filtering for transients and for intervals with bad field-of-view (which impacts the velocity measurement) was done in the same way here as in the main text.



\section{Sensitivity of the Stochastic Heating Rate to Shape of the VDF} \label{app:sh-vdf}

To address how the shape of the 1d velocity distribution function, $f_\perp(v_\perp)$, impacts the stochastic heating (SH) rate in Equation~\ref{eqn:Qsh}, we compare the results of Figure~\ref{fig:sh-let}(b) with calculations that utilize 1d distribution functions with flatter cores. We define:
\begin{equation}
    f_\perp = C e^{-(v_\perp / a)^p} 
\end{equation} where $C$ and $a$ are determined from physical constraints and $a$ is a function of $v_{th, \perp}$. 

$C$, is determined from requiring the zeroth order moment of the distribution function to equal the measured QTN density ($n_p$): 
\begin{equation}
    n_p = \int_0^\infty 2 \pi v_\perp f_\perp d v_\perp =   \int_0^\infty 2 \pi v_\perp C e^{-(v_\perp / a)^p} d v_\perp 
\end{equation}
where we define $u = (v_\perp / a)^p$ giving $v_\perp = a u^{1/p}$ and $d v_\perp = \frac{a}{p} u^{1/p -1}$. We then can solve for $C$:
\begin{equation}
      n_p =  2 \pi C \int_0^\infty a u^{1/p} e^{-u} \frac{a}{p} u^{1/p - 1} d u = 2 \pi C \frac{a^2}{p} \int_0^\infty u^{2/p - 1} e^{-u} du = 2 \pi C \frac{a^2}{p} \Gamma(2/p)
\end{equation}
to get 
\begin{equation}
    C = \frac{n_p p}{2 \pi a^2 \Gamma(2/p)}.  
\end{equation}

$a$ is determined by requiring the second order moment, to match the measured temperature: 
\begin{multline}
    v_{th, \perp}^2 = \frac{1}{n_p} \int_0^\infty v_\perp^2 2 \pi v_\perp f_\perp (v_\perp) d v_\perp = 2 \pi C \int_0^\infty v_\perp^3 e^{-(v_\perp / a)^p} d v_\perp \\
     = \frac{2 \pi C}{n_p}  \int_0^\infty  a^3 u^{3/p} e^{-u} \frac{a}{p} u^{1/p - 1} d u =  \frac{2 \pi C}{n_p} \frac{a^4}{p} \int_0^\infty u^{3/p} e^{-u} \frac{a}{p} u^{1/p - 1} d u =  \frac{2 \pi C}{n_p} \frac{a^4}{p} \Gamma(4/p)
\end{multline}
which then determines $a$ as:
\begin{equation}
    a = v_{th, \perp} \sqrt{\frac{\Gamma(2/p)}{\Gamma(4/p)}}.
\end{equation}

\begin{figure}[ht!]
    \centering
    \includegraphics[width=0.8\linewidth]{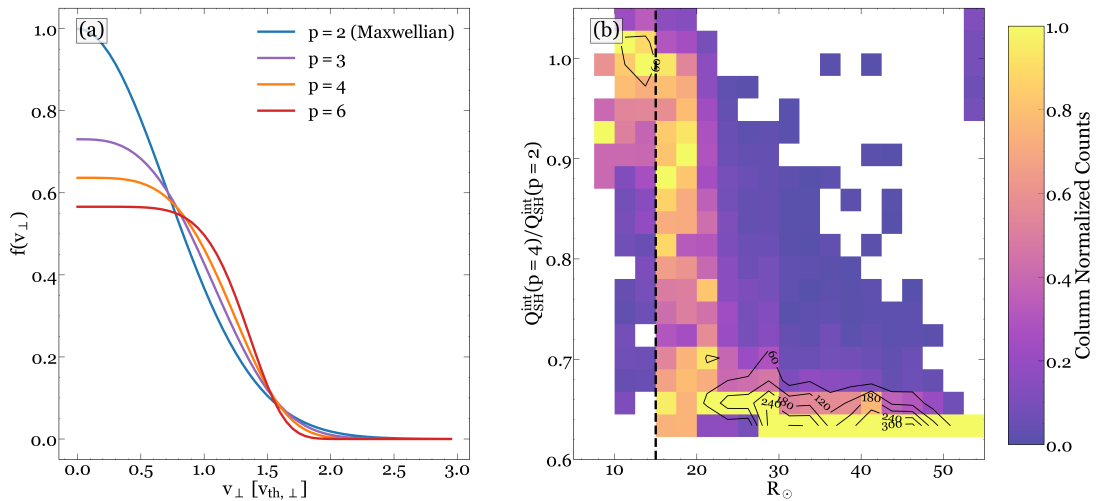}
    \caption{(a): Cartoon example of VDF shapes for different values of $p$. (b): Ratio of the intermittent stochastic heating rate for the $p=4$ case with the Maxwellian ($p=2$) case as a function of heliocentric distance ($R_\odot$).}
    \label{fig:gauss-vdf}
\end{figure}

This reduces to a 1d Maxwellian distribution function when $p = 2$, and is flatter at low $v_\perp$ when $p > 2$ as shown in Figure~\ref{fig:gauss-vdf}(a), similar to the modified Moyal distribution shown by \citet{Klein-2016} to be produced by SH. We find that the flattening of the VDF does not change the main result of the manuscript.

Figure~\ref{fig:gauss-vdf}(b) compares the ratio of the intermittent SH rate as a function of heliocentric distance for $p=4$ with the Maxwellian case ($p=2$) shown in the main text (Section~\ref{sec:res}). The $p=3$ and $p=6$ cases show equivalent distance dependent results. Closer to the Sun, in the region where we show SH to be the relevant mechanism, we find that the flattening of the VDF does not significantly suppress the SH rate for any of the test cases. In the region between 15 and 20 $R_\odot$, where we show the SH rate to drop off compared with the LET rate (see Figure~\ref{fig:sh-let}), flattening the VDF starts to have a larger impact on the SH rate. At distances farther from the Sun, flattening the VDF decreases the SH rate to $\sim$75\% (p=3),  $\sim$65\% (p=4), $\sim$60\% ($p=6$) of the Maxwellian result. This is approximately equivalent to the decrease in amplitude at the lowest perpendicular velocities between the Maxwellian VDF and the “flattened” VDF.

\end{document}